\documentclass[12pt]{article}

\usepackage[utf8]{inputenc}
\usepackage{graphicx}
\usepackage{etoolbox}
\usepackage{hyperref}
\usepackage{times}
\usepackage{bm}
\usepackage{makecell}
\usepackage{color}
\usepackage{xcolor}
\usepackage{amssymb}
\usepackage{amsmath}
\usepackage{braket}
\usepackage{siunitx}
\usepackage{lineno}
\usepackage{caption}
\usepackage{cite}  

\newenvironment{sciabstract}{%
\begin{quote} \bf}
{\end{quote}}

\graphicspath{ {figures/} }

\makeatletter
\let\saved@includegraphics\includegraphics
\AtBeginDocument{\let\includegraphics\saved@includegraphics}
\makeatother

\DeclareUnicodeCharacter{2212}{-}

\begin{document}

\title{Pressure‑Induced 18 K Superconductivity and Two Superconducting Phases in CuIr$_2$S$_4$}

\date{}


\author{%
  \begin{minipage}{\textwidth}
    \centering
    Bijuan Chen$^{1*}$, Yuhao Gu$^{2}$, Dong Wang$^{1}$, Dexi Shao$^{2,9}$, Wen Deng$^{1}$, Xin Han$^{2}$, Meiling Jin$^{3}$, Jing Song$^{2}$, Yu Zeng$^{2}$, Hirofumi Ishii$^{4}$, Yen-Fa Liao$^{4}$, Dongzhou Zhang$^{5}$, Jianbo Zhang$^{1}$, Youwen Long$^{2,8}$, Jinlong Zhu$^{3}$, Liuxiang Yang$^{1}$, Hong Xiao$^{1}$, Jia-cai Nei$^{6}$, Youguo Shi$^{2}$, Changqing Jin$^{2}$, Jiangping Hu$^{2*}$, Ho-kwang Mao$^{1,7}$, Yang Ding$^{1*}$\\[1.5ex]
    {\normalsize $^{1}$ Center for High Pressure Science \& Technology Advanced Research, Beijing, 100094, China}\\
    {\normalsize $^{2}$ Institute of Physics, Chinese Academy of Sciences, Beijing 100190, China}\\
    {\normalsize $^{3}$ Department of Physics and Shenzhen Engineering Research Center for Frontier Materials Synthesis at High Pressures, Southern University of Science and Technology (SUSTech), Shenzhen 518055, China}\\
    {\normalsize $^{4}$ National Synchrotron Radiation Research Center, Hsinchu 30076, Taiwan}\\
    {\normalsize $^{5}$ Hawaii Institute of Geophysics \& Planetology, University of Hawaii Manoa, Honolulu, HI 96822, USA}\\
    {\normalsize $^{6}$ Department of Physics, Beijing Normal University, Beijing 100875, China}\\
    {\normalsize $^{7}$ Shanghai Advanced Research in Physical Sciences (SHARPS), Shanghai 201208, China}\\
    {\normalsize $^{8}$ Songshan Lake Materials Laboratory, Dongguan, Guangdong 523808, China}\\
    {\normalsize $^{9}$ Department of Physics, Hangzhou Normal University, Hangzhou 311121, China}\\[1ex]
    {\normalsize $^*$To whom correspondence should be addressed; E-mail: bijuanchen.5459@gmail.com; jphu@iphy.ac.cn; yang.ding@hpstar.ac.cn}
  \end{minipage}
}

\date{}

\baselineskip24pt
\maketitle
\newpage
\vspace{5mm}
\begin{sciabstract}

We report pressure-induced superconductivity in the spinel CuIr$_{2}$S$_{4}$ with a transition temperature ($T_{\text{c}}$) reaching \textbf{18.2 K}, establishing a new record for this class of materials and surpassing the decades-old limit of 13.7 K. Our electrical transport and synchrotron X-ray diffraction studies up to 224 GPa reveal the emergence of \textbf{two distinct superconducting phases} from a charge-ordered insulating state. The first phase (SC-I) appears around 18 GPa, and forms a dome-shaped superconducting region in which the resistivity exhibits a pronounced, field- and current-sensitive drop without reaching strict zero above our base temperature. Above 111.8 GPa, a second, lower-$T_{\text{c}}$ phase (SC-II) emerges and coexists with SC-I over a broad pressure range, and SC-II ultimately develops a true zero-resistance state above 122.2 GPa. These superconducting phases are intimately linked to a cascade of structural transitions that systematically distort the frustrated pyrochlore lattice of Ir atoms. Our results expand the potential for superconductivity in spinels and demonstrate a pathway to high-$T_{\text{c}}$ pairing directly from a correlated insulating state driven by lattice tuning.

\end{sciabstract}

\maketitle

\newpage

Superconductivity in geometrically frustrated materials provides a unique platform to investigate unconventional pairing mechanisms and achieve higher superconducting transition temperatures ($T_c$). Spinel compounds (AB$_2$X$_4$, where A and B are transition metals and X is a chalcogen or oxygen) have attracted significant interest due to their rich electronic phenomena arising from the complex interplay between spin, charge, and orbital degrees of freedom\cite{Patoux2009High,Vestal2003Synthesis,Senn2011Charge,Liu2000Chemical,Lee2002Emergent}. This interplay often leads to diverse emergent behaviors\cite{Jin2015Anomalous,Radaelli2002Formation,Urano2000LiV2O4,Verwey1939Electronic}, including metal--insulator transitions (MIT)\cite{Furubayashi1994Structural,Ma2017Opposite}, magnetic ordering\cite{Verwey1939Electronic}, superconductivity\cite{Ito2003Pressure,Moshopoulou1999Superconductivity}, among other correlated phenomena. Despite extensive research efforts, the maximum of $T_c$ in spinel superconductors has remained around 13.7~K for several decades, exemplified by LiTi$_2$O$_4$\cite{Johnston1976Superconducting,Okada2017Scanning}. More recently, superconductivity with $T_c$ up to 16 K was reported in thin surface layers of V-doped MgTi$_2$O$_4$~\cite{rahaman2023surface}; however, this superconductivity only emerges in thin films and has not been realized in bulk form. Overcoming this longstanding limit in bulk spinel materials remains both a fundamental scientific challenge and an opportunity to discover novel pathways toward higher-temperature superconductivity.

\begin{figure}[ht]
    \centering
    \includegraphics[width=1.0\linewidth]{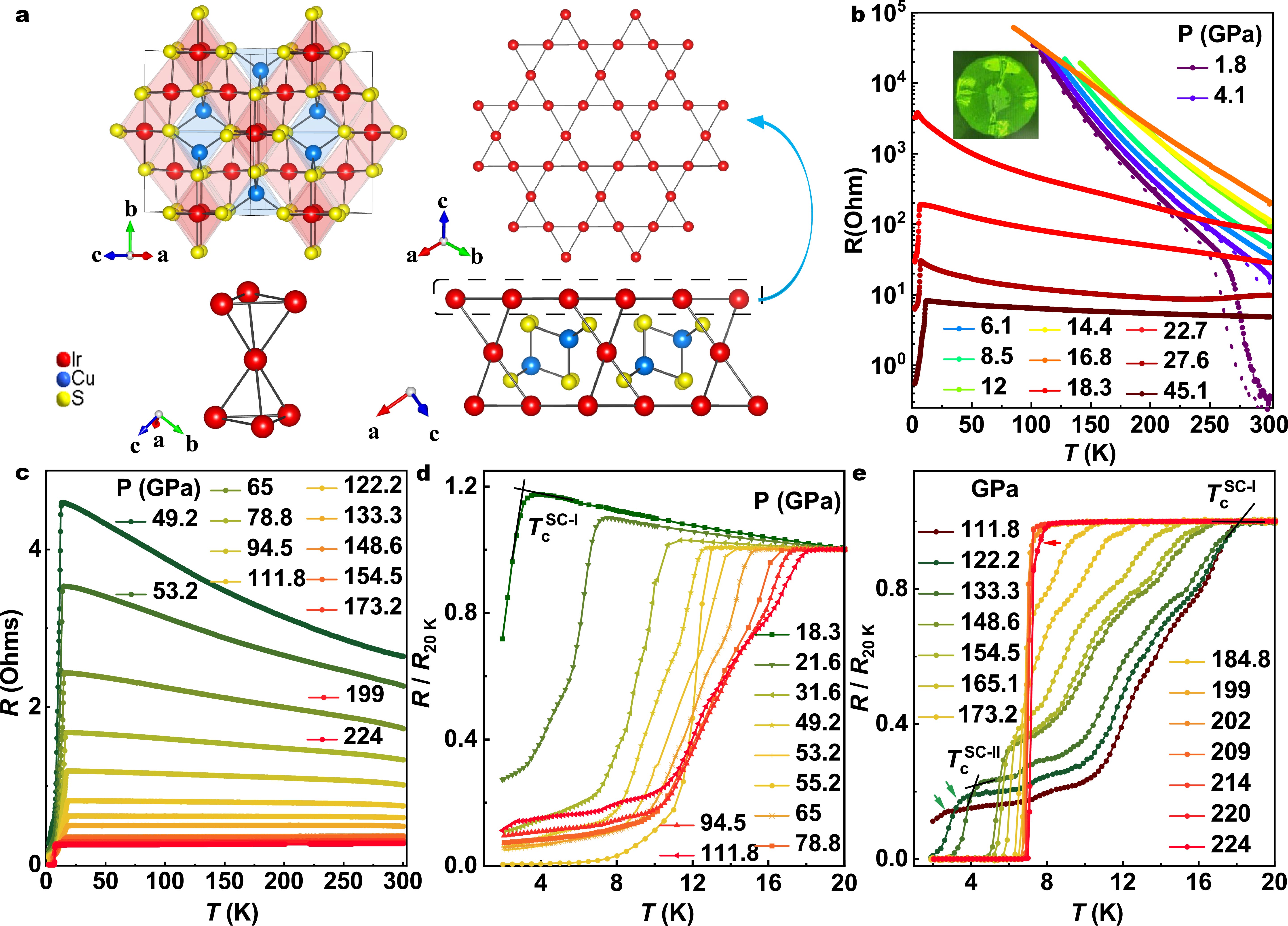}
      \caption{\textbf{Pressure‑induced superconductivity in CuIr$_2$S$_4$.} (\textbf{a}) The crystal structure of CuIr$_2$S$_4$, space group \textit{Fd-3m}. The Ir sublattice forms a network of corner‑sharing tetrahedra and a perfect 2D Kagome‑like net~\cite{Ortiz2019New}. (\textbf{b}, \textbf{c}) Temperature dependence of resistance of CuIr$_2$S$_4$ under various pressures up to 224\,GPa. (\textbf{d}, \textbf{e}) An expanded view of the low‑temperature resistive region under pressures ranging from 18.1–111.8\,GPa (\textbf{d}) and 122.2–224\,GPa (\textbf{e}). At 111.8\,GPa, a new superconducting phase emerges (marked with green arrow) and is enhanced with increasing pressure. The criterion for determining the superconducting transition temperature ($T_{\rm c}$) is shown with a black cross‑wire. $T_c$ of SC-I and SC-II here is defined as the onset of the resistance drop.}
  \label{fig:1}
\end{figure}

Within this context, CuIr$_2$S$_4$ stands out due to its distinctive electronic and structural properties. At ambient pressure, CuIr$_2$S$_4$  adopts a cubic spinel structure, characterized by a pyrochlore lattice of corner-sharing tetrahedra formed by Ir atoms occupying the B-sites (Fig. ~\ref{fig:1}a). This lattice arrangement comprises alternating layers of Kagome-like and triangular structures stacked along the [111] crystallographic direction~\cite{Ortiz2019New}. This configuration has gained significant interest due to its intrinsic geometric frustration, which leads to highly degenerate electronic states and the emergence of exotic correlated phases, including unconventional superconductivity, quantum spin liquids, and charge-ordered states, etc.

CuIr$_2$S$_4$ experiences a pronounced MIT around 230~K at ambient conditions, accompanied by a structural distortion from cubic to triclinic symmetry driven by charge ordering of the Ir ions\cite{Radaelli2002Formation,Nagata1994Metal}. This insulating state suppresses superconductivity under normal conditions. In contrast, closely related spinel compounds such as CuRh$_2$S$_4$ and CuRh$_2$Se$_4$ do not undergo MIT or structural transitions, yet they exhibit superconductivity\cite{Ito2003Pressure,Robbins1967Superconductivity}, indicating that subtle structural and electronic variations significantly influence the emergence of superconducting phenomena. Additionally, Zn doping in CuIr$_2$S$_4$ partially suppresses the MIT and induces superconductivity\cite{Suzuki1999Metal}, albeit with relatively low $T_c$ values (2--3~K), underscoring the delicate balance between structural integrity, charge order, and superconductivity.

Although chemical substitution offers valuable insights, it inherently introduces disorder, limiting achievable superconducting temperatures and motivating cleaner tuning methods. High pressure emerges as an ideal alternative, providing a disorder-free pathway to substantially modify lattice parameters and electronic bandwidths, potentially drive the transition from insulating to superconducting states\cite{Chen2019Probing,Chen2022Novel,Mao2018Solids}.

While prior theoretical work predicted that pressure could induce superconductivity in CuIr$_2$S$_4$, experimental validation has remained elusive, with modest pressures only serving to enhance the insulating state\cite{clay2019charge,Oomi1995Effect,garg2007reentrant,bovzin2014cu}.

In this letter, we systematically explore CuIr$_2$S$_4$ under high pressures up to 224~GPa using comprehensive electrical transport, synchrotron X-ray diffraction and Raman spectroscopy. Our results reveal the suppression of the ambient insulating phase and uncover superconductivity hidden beneath. Remarkably, we observe two distinct superconducting phases: an initial phase (SC-I) emerging at 3.8~K, steadily increasing to a record-high $T_c$ of 18.2~K at 133~GPa— exceeding the longstanding $T_c$ limit for spinel superconductors and establishing a new benchmark for iridium-based superconductors. Additionally, a second, lower-temperature superconducting phase (SC-II) emerges at pressures above approximately 111.8~GPa, revealing unprecedented complexity in the superconducting phase landscape of spinel compounds. Our findings mark a significant advancement in understanding superconductivity in geometrically frustrated materials, highlighting the powerful role of extreme pressure for engineering novel electronic states. This work not only redefines the performance limits of spinel superconductors but also provides essential insights into exploring and designing future high-temperature superconductors in correlated electron systems.

To probe pressure-induced electronic transitions, we conducted comprehensive resistance measurements at pressures from 1.8 to 224 GPa using diamond anvil cells with various culet sizes. With increasing pressure, the MIT temperature (\textit{T}\textsubscript{M}) shifts upward, completely suppressing metallic conductivity above 4.1 GPa (Fig. 1b and Fig. S1~\cite{SM}).\nocite{Matsumoto2000Single,Bruker2009APEX,Sheldrick2015SHELXT,Prescher2015DIOPTAS,Toby2013GSAS,errandonea2006high,li1987measurement,chen2012high,Kresse1996Efficiency,Kresse1999From,Perdew1996Generalized,ouahrani2023effect,Barzykin2002Inhomogeneous,Liu2021Two,Jin2021Discovery}
 Between 4.1 and 16.8\,GPa, CuIr$_2$S$_4$ exhibits predominantly insulating behavior, as indicated by a negative temperature coefficient of resistance. 

Remarkably, at approximately 18 GPa, a superconducting-like transition appears (phase SC-I), characterized by a pronounced resistance drop around  $T_c \approx$ 3.8\,K (Figs.~\ref{fig:1}b and 1d).
Here, $T_c$ is defined by the onset of this resistance drop. As pressure increases, the superconducting-like feature becomes more pronounced and the residual resistance decreases, with nearly zero resistance first observed around 53.2 GPa [Fig. 1(d)]. Upon further compression, however, a pronounced low-temperature tail develops again, coincident with the pressure range where the lower-$T_c$ SC-II phase begins to emerge. Such a substantial resistance drop that does not reach strict zero is common in high-pressure experiments~\cite{sun2023signatures,mandyam2025uncovering} and is usually interpreted as filamentary superconductivity or granularity arising from non-hydrostatic conditions and phase coexistence. In CuIr$_2$S$_4$, the SC-I regime shows analogous behavior, consistent with percolative superconducting paths embedded in a competing charge-ordered matrix, as seen in other high-pressure systems with coexisting phases. It is clearly seen that the evolution of $T_c$ under increasing pressure is non-monotonic: initially, $T_c$ increases from 3.8 to 16 K as pressure increases from 18 to 78.8 GPa, then it slows down and peaks at 18.2 K at 133.3 GPa, before decreasing monotonically with further pressure increase.

Surprisingly, at 111.8 GPa, a second superconducting phase (SC-II) is observed, demonstrated by a resistance drop at approximately 2.2 K, marked by a green down-pointing arrow in Fig.~\ref{fig:1}e. This resistance drop becomes more pronounced with additional compression, ultimately reaching zero resistance at 122.2 GPa. The observed two-step superconducting transition parallels phenomena previously documented in materials such as Bi$_2$Sr$_2$CaCu$_2$O$_{8+\delta}$ \cite{Guo2019Crossover} and ZrTe$_5$\cite{Zhou2016Pressure}, highlighting the complex interplay and rich diversity of pressure-driven superconductivity in CuIr$_2$S$_4$. Notably, CuIr$_2$S$_4$ exhibits a rare transition directly from an insulating state to a superconducting phase, bypassing the intermediate metallic state typically observed in other systems. This sets it apart from most known spinel-derived superconductors—such as the lacunar (cluster) spinels GaM$_4$X$_8$ (M = Nb, Ta; X = S, Se)—which generally follow a conventional insulator–metal–superconductor sequence~\cite{deng2021metallization,pocha2005crystal}.

\begin{figure}[ht]
  \centering
  \includegraphics[width=0.95\linewidth]{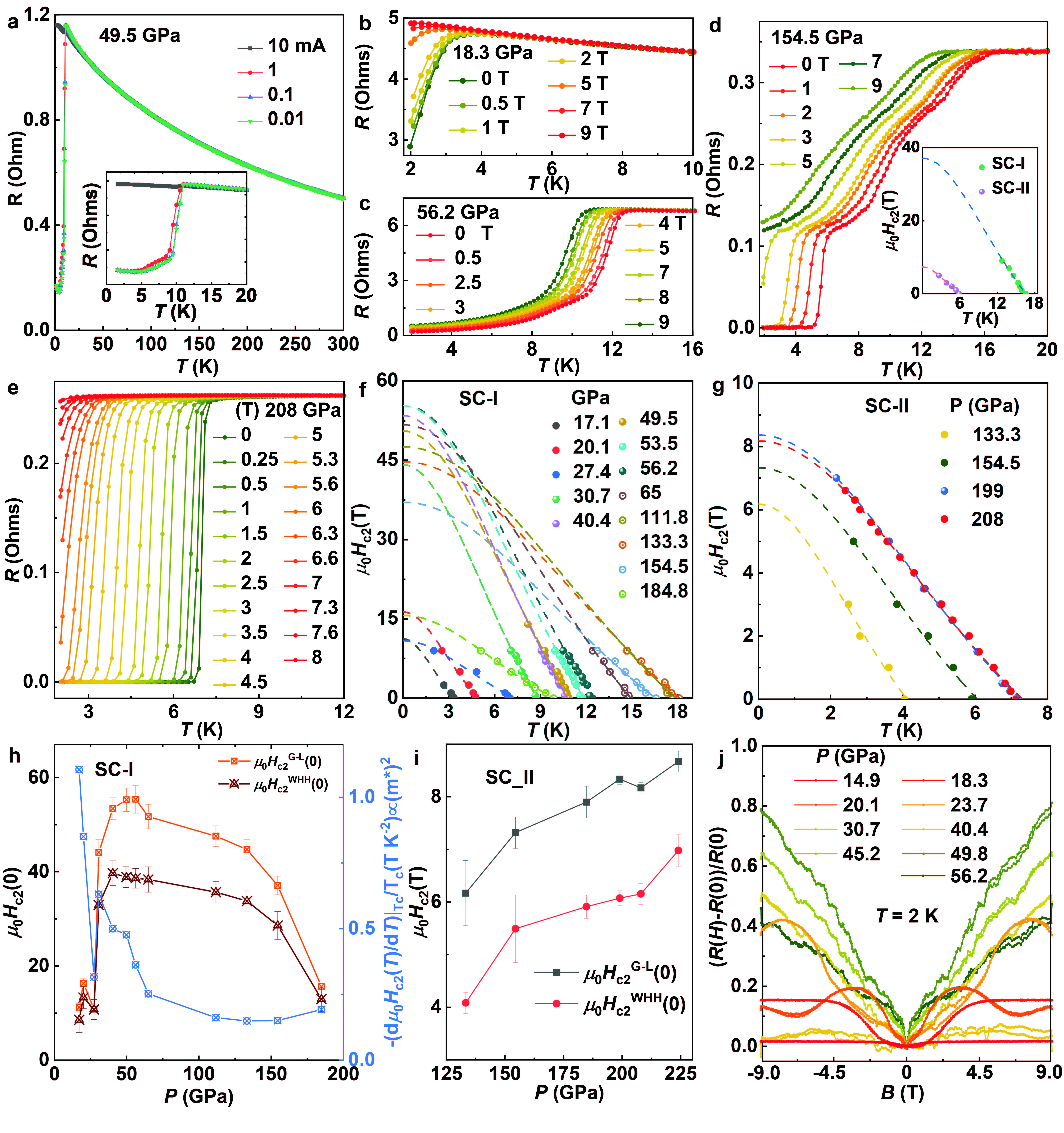}
  \caption{\textbf{Evolution of the superconducting transition in CuIr$_2$S$_4$ under magnetic fields.} 
(\textbf{a}) Temperature dependence of the electrical resistance at 49.5\,GPa under various applied currents. Inset: enlarged view of $R(T)$ at 49.5\,GPa under different currents. 
(\textbf{b}, \textbf{c}) Temperature-dependent resistivity under varying magnetic fields at 18.3\,GPa and 56.2\,GPa, respectively. 
(\textbf{d}, \textbf{e}) Temperature-dependent resistivity under different magnetic fields at 154.5\,GPa and 208\,GPa, respectively. Inset of (\textbf{d}) shows the best fit of $T_{\rm c}$ vs.\ $\mu_0 H_{c2}(0)$ using the G–L formula [Eq.~(1)]. 
(\textbf{f}, \textbf{g}) Fits of $T_{\rm c}$ vs.\ $\mu_0 H_{c2}(0)$ using the G–L formula for the SC-I and SC-II phases, respectively. 
(\textbf{h}, \textbf{i}) Pressure dependence of the zero-temperature upper critical field $\mu_0H_{c2}(0)$ obtained from the G–L fit [Eq.~(1)] and the WHH model [Eq.~(2)], along with the normalized slope $-\left( \frac{1}{T_c} \left. \left[ \frac{d\mu_0 H_{c2}(T)}{dT} \right] \right|_{T_c} \right)$. Error bars for $\mu_0 H_{c2}(0)$ are derived from the covariance matrix of the Ginzburg--Landau fit and are typically $\pm 1$--3~T. (\textbf{j}) Magnetic field dependence of the resistance measured at $T = 2$\,K under various pressures.}

  \label{fig:2}
\end{figure}

To further characterize the superconducting behavior of CuIr$_2$S$_4$ single crystals under varying pressures, we systematically investigated the influence of electrical current and magnetic fields on resistance. As depicted in Fig.~\ref{fig:2}a, increasing the applied current notably suppresses the resistance drop, indicative of superconducting onset, with a critical current estimated at approximately 10~mA. Figs.~\ref{fig:2}b and 2c display $R(T)$ curves at pressure of 18.3 and 56.2~GPa, respectively, under various magnetic fields, where only the SC-I phase is observed. In both cases, the superconducting transition temperature $T_c$ decreases monotonically with increasing magnetic field. Notably, at 154.5~GPa, the $R(T)$ curves bifurcate into two distinct segments (Fig.~\ref{fig:2}d) under different magnetic fields, signifying the coexistence of SC-I and SC-II. SC-II exhibits heightened sensitivity to magnetic fields; its superconducting transition becomes indiscernible above 5~T, leaving SC-I as the dominant phase. At a higher pressure of 208~GPa, only SC-II persists (Fig.~\ref{fig:2}e).

To quantify the upper critical field at zero temperature, $\mu_0 H_{c2}(0)$, we employed the empirical Ginzburg--Landau (G--L) model\cite{Li2022Pressure,Wang2022Pressure}:
\begin{equation}
\mu_0 H_{c2}(T) = \mu_0 H_{c2}(0) \frac{1 - t^2}{1 + t^2},
\end{equation}
where $t = T/T_c$. At 154.5~GPa, $\mu_0 H_{c2}(0)$ is estimated to be 37.0~T for SC-I and 7.3~T for SC-II (inset, Fig.~\ref{fig:2}d). The significant difference in the estimated values of $\mu_0 H_{c2}(0)$ between SC-I and SC-II underscores their fundamentally distinct superconducting characteristics.

Intriguingly, $\mu_0 H_{c2}(0)$ for SC-II increases monotonically with pressure, paralleling the enhancement of $T_c^{\mathrm{SC-II}}$ (Fig.~\ref{fig:2}g and 2i). In contrast, SC-I exhibits a non-monotonic pressure dependence (Fig.~\ref{fig:2}h): $\mu_0 H_{c2}(0)^{\mathrm{SC-I}}$ initially increases from $\sim$11.1~T at 18.1~GPa to 16.3~T at 20.1~GPa, then decreases to $\sim$10.9~T at 27.4~GPa. Beyond this, it rises sharply to 44.1~T at 30.7~GPa and continues to grow gradually to $\sim$55.4~T at 56.2~GPa. Upon further compression, it decreases to 47.6~T at 111.8~GPa and drops steeply to 15.7~T at 184~GPa, potentially indicating suppression of the SC-I phase. 

To corroborate these results, we also applied the Werthamer--Helfand--Hohenberg (WHH) model\cite{Li2022Superconductivity}:
\begin{equation}
\mu_0 H_{c2}^{\mathrm{WHH}}(0) = -0.69\, T_c \left. \frac{d H_{c2}}{dT} \right|_{T=T_c},
\end{equation}
based on the initial slope near $T_c$. The obtained values, depicted in Fig.~\ref{fig:2}h and 2i, yield slightly lower $\mu_0 H_{c2}(0)$ values but follow the same pressure-dependent trend observed with the G--L model. In the single-band framework, the initial slope near $T_c$ is expected to be inversely proportional to the square of the Fermi velocity $v_F$ and directly proportional to the effective mass $m^*$ of the charge carriers\cite{Kaluarachchi2016Nonmonotonic,Kogan2012Orbital}:
\begin{equation}
-\left( \frac{1}{T_c} \left. \left[ \frac{d\mu_0 H_{c2}(T)}{dT} \right] \right|_{T_c} \right)
 \propto v_F^{-2} \propto m^*
\end{equation}

We further extract the normalized slope $-\left( \frac{1}{T_c} \left. \left[ \frac{d\mu_0 H_{c2}(T)}{dT} \right] \right|_{T_c} \right)
$, which decreases from $\sim$1.10~T/K$^2$ at 18.1~GPa to $\sim$0.19~T/K$^2$ at 184.8~GPa (Fig.~\ref{fig:2}h), suggesting a pressure-induced reduction in $m^*$ or electronic correlations, likely associated with bandwidth broadening under pressure\cite{Wang2022Pressure}. A sudden increase around 27~GPa may reflect an underlying structural transition.

Further insights are obtained from magnetoresistance ($MR$) measurements in the SC-I phase (Fig.~\ref{fig:2}j) at 2~K under various pressures. Pronounced magnetoresistance with oscillation-like features appears between $\sim 18$ and $27$ GPa and is strongly reduced above $\sim 30$ GPa [Fig. 2(j)]. These changes indicate substantial modifications of the electronic structure—whether via Fermi-surface reconstruction, enhanced scattering, or magnetic correlations—in proximity to a pressure-driven transition. Because our measurements are limited to transport, we cannot distinguish uniquely between these possibilities, and we therefore refer to them as signatures of a nearby quantum phase transition rather than direct evidence for magnetic order. Notably, the suppression of these magnetoresistance oscillations coincides with the emergence of structural Phase V around 27 GPa, suggesting a coupling between the electronic instability and the subtle reorganization of the orthorhombic lattice.

\begin{figure}[ht]
  \centering
  \includegraphics[width=\linewidth]{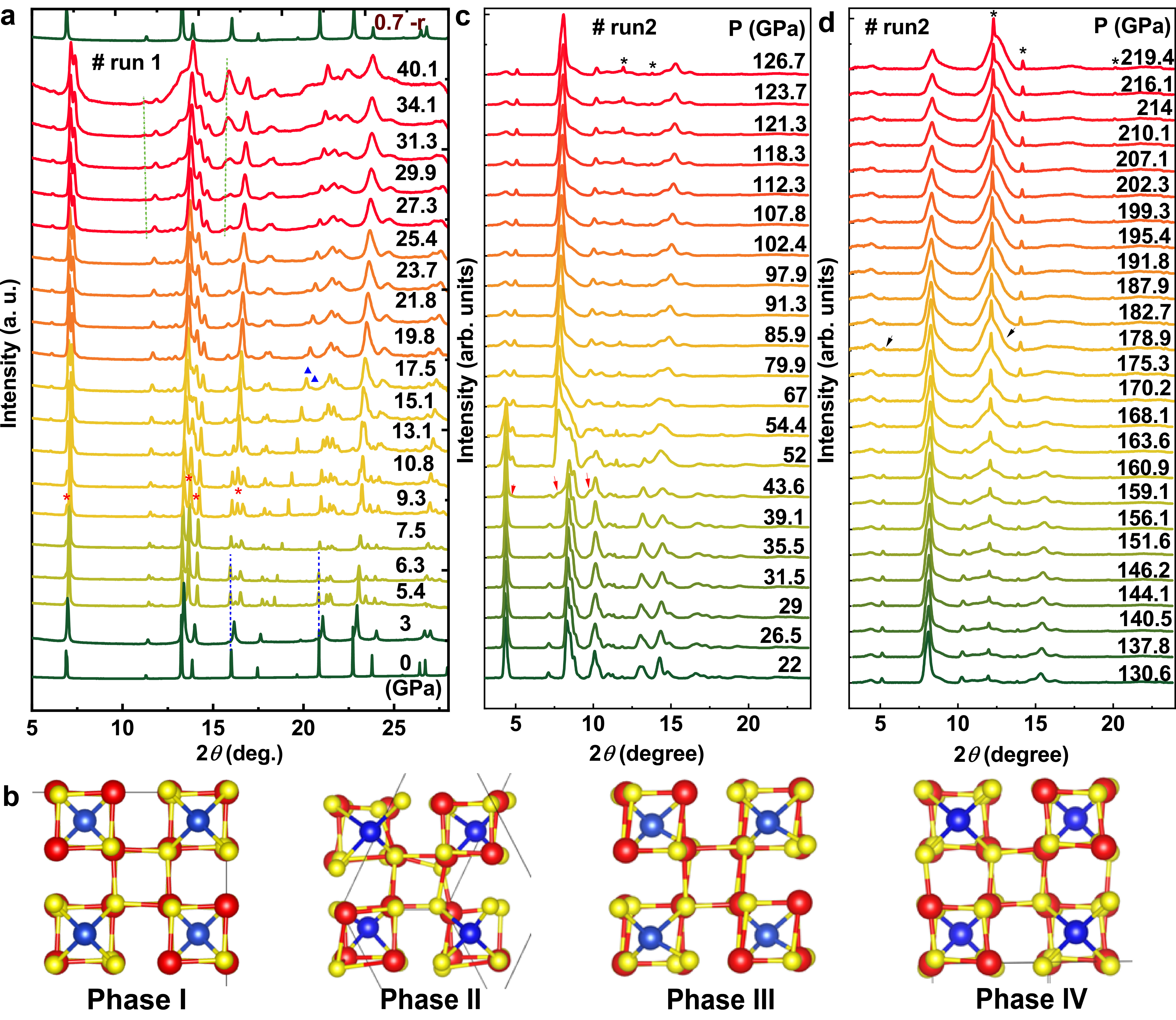}
  \caption{\textbf{Pressure‑induced structural phase transition in CuIr$_2$S$_4$.} (\textbf{a}) XRD patterns measured in CuIr$_2$S$_4$ under high pressure up to 40.1\,GPa with an incident wavelength $\lambda = 0.68883\,\text{\AA}$. Asterisks, triangles and dotted lines indicate the presence of extra peaks. (\textbf{b}) Pressure‑induced transformation of crystalline from metallic cubic phase to triclinic insulator phase and finally orthorhombic superconductive phase. (\textbf{c}, \textbf{d}) Synchrotron XRD patterns with subtracted background of CuIr$_2$S$_4$ at selected pressures up to 219.4\,GPa and incident wavelength $\lambda = 0.434\,\text{\AA}$. Arrows indicate the presence of extra peaks. Black asterisks indicate the Au peaks for pressure calibration.}
  \label{fig:4}
\end{figure}

Having identified the emergence of superconductivity in transport measurements, we next sought to uncover the structural evolution that may drive this transition. To this end, we performed high-pressure powder X-ray diffraction (PXRD) measurements on CuIr$_2$S$_4$ up to 219.4 GPa at room temperature, as shown in Fig.~\ref{fig:4}. In Fig.~\ref{fig:4}a, at ambient conditions, CuIr$_2$S$_4$ crystallizes in the cubic spinel structure (space group $Fd\bar{3}m$) with lattice parameter $a \approx 9.8677\,\text{\AA}$ (detailed in Table SI)~\cite{Furubayashi1994Structural,Radaelli2002Formation}.  Upon application of pressure, the diffraction peaks shift systematically to higher $2\theta$ angles, indicating continuous shrinkage of the lattice.  The first clear structural change occurs around 3\,GPa: new Bragg peaks appear (marked by blue dotted lines in the XRD patterns), signaling the onset of a lower-symmetry phase.  By 5.4\,GPa this phase is fully developed, and Rietveld refinements identify it as a triclinic structure (space group $P\bar{1}$, as shown in Fig. S2).  We label this high-pressure triclinic phase as Phase~II (the ambient cubic phase being Phase~I).  The triclinic Phase~II is characterized by the same kind of charge ordering (Ir$^{3+}$/Ir$^{4+}$) and spin dimerization of Ir ions that occur in the ambient-pressure insulating state (the low-temperature triclinic phase of CuIr$_2$S$_4$\cite{Radaelli2002Formation}).  In other words, applying a modest pressure of a few gigapascals stabilizes the dimerized insulating state even at room temperature, consistent with the rapid suppression of the MIT by pressure noted above. With further compression, additional phase transitions take place.  Phase III (another triclinic modification) appears at $\sim$9.3\,GPa, indicated by changes in the diffraction pattern (red asterisks), and is followed by another transition to Phase~IV at $\sim$17.5\,GPa (blue triangles).  Phase IV was identified via single-crystal XRD as an orthorhombic structure (Fig. S2), marking a transition from the triclinic lattice of Phase~II/III to a more symmetric orthorhombic lattice.  Notably, the emergence of Phase~IV at $\sim$17.5\,GPa coincides with the pressure range where superconductivity first appears (SC‑I).  This may indicate that the orthorhombic Phase~IV provides a favorable structural environment for the onset of superconductivity. 

From a chemical coordination standpoint, the crystal structure of CuIr$_2$S$_4$ comprises distinct [IrS$_6$] octahedra and [CuS$_4$] tetrahedra, which serve as the primary units for analyzing structural evolution under pressure, as illustrated in Fig.~S3. We find that the most significant structural changes under pressure involve the [IrS$_6$] octahedral framework, as illustrated in Fig.~\ref{fig:4}b. The [IrS$_6$] octahedra become increasingly distorted as pressure rises, whereas the [CuS$_4$] tetrahedra are relatively rigid in comparison.  This differential response is evident in our refinements (see Supplementary Fig.~S2) and reflects the different compressibilities and bonding of the two sublattices. The rapid increase in octahedral distortion correlates with each structural phase change, underscoring that the Ir–Ir distances and Ir–S–Ir bond angles are being tuned by pressure into configurations that dramatically alter the electronic ground state. These changes reflect the intricate ways in which CuIr$_2$S$_4$ relieves the internal strain and frustration of the Ir–S network under pressure.

Remarkably, the pressure-induced superconductivity in CuIr$_2$S$_4$ is accompanied by—and evidently intertwined with—these structural transformations. This contrasts with many other superconducting spinels (e.g.\ LiTi$_2$O$_4$, CuRh$_2$S$_4$) where superconductivity occurs in the original cubic phase without any symmetry lowering\cite{Ito2003Pressure,Moshopoulou1999Superconductivity,Robbins1967Superconductivity,Suzuki1999Metal, Hagino1995Superconductivity}. In CuIr$_2$S$_4$, however, the onset of superconductivity (SC‑I) is associated with leaving the cubic phase and entering an orthorhombic phase (Phase~IV), suggesting a strong coupling between lattice distortions and superconductivity.  

As pressure increases further into the tens of gigapascals, we detect yet another structural modification: two extra diffraction peaks emerge around 11.2° and 15.7° ($2\theta$) at pressures above $\sim$27.3\,GPa (indicated by the green dashed line in Fig.~\ref{fig:4}a). These extra peaks intensify with continued compression, indicating the development of a Phase V.  The appearance of Phase V around 27\,GPa is particularly intriguing because it coincides with an enhancement of the superconducting $\mu_0 H_{c2}(0)$ (noted above in Fig.~\ref{fig:2}h). This Phase~V may represent a subtle reorganization of the orthorhombic structure or a transition to another structure altogether; in either case, its concurrence with changes in superconducting behavior highlights a strong coupling between lattice and superconductivity in CuIr$_2$S$_4$. The sequential Phase~I$\to$ Phase~II $\to$ Phase~III $\to$ Phase~IV $\to$ Phase~V transitions are also corroborated by Raman spectroscopy, as detailed in Fig.~S4 \cite{Zwinscher1995Lattice,Ammundsen1999Lattice,Zhang2010Study,Naseska2020Orbitally}.
  
Additionally, another structural phase transition emerges at 43.6 GPa (phase VI), identified by the appearance of new Bragg peaks (indicated by the red arrow in Fig.~\ref{fig:4}c), which completes transformation around 67 GPa. Phase VI is notably robust, showing no further changes up to 178.9 GPa. Above 178.9 GPa, a phase transition into phase VII is evidenced by the disappearance of a peak around 5.2° and the concurrent appearance of a broad peak near 12.6° in Fig.~\ref{fig:4}d (highlighted by black arrow). Beyond this phase, X-ray diffraction patterns remain unchanged up to the highest studied pressure of 219 GPa, indicating no further significant symmetry alterations; thus, the structure likely maintains the phase VII configuration throughout the remainder of the pressure range. Notably, this final high-pressure structural phase transition aligns closely with the pressure at which SC-I transitions to SC-II (approximately 184 GPa). Furthermore, no additional diffraction peaks attributable to sulfur are detected\cite{Akahama1993Pressure,Luo1993Beta}, confirming that all structural transitions are intrinsic to CuIr$_2$S$_4$ and establishing that the observed superconductivity is solely associated with CuIr$_2$S$_4$.

This crystallographic study reveals a rich progression of phases: from the initial cubic spinel (Phase I) to a dimerized triclinic insulator (Phase II), followed by an intermediate Phase III, an orthorhombic Phase IV, and three additional high-pressure phases (Phase V to VII). These structural phases underpin and contextualize the electronic phases observed in transport measurements, firmly establishing that pressure-driven lattice rearrangements are one of the most critical factors responsible for the dramatic superconducting phenomena in CuIr$_2$S$_4$.

\begin{figure}[ht]
  \centering
   \includegraphics[width=\linewidth]{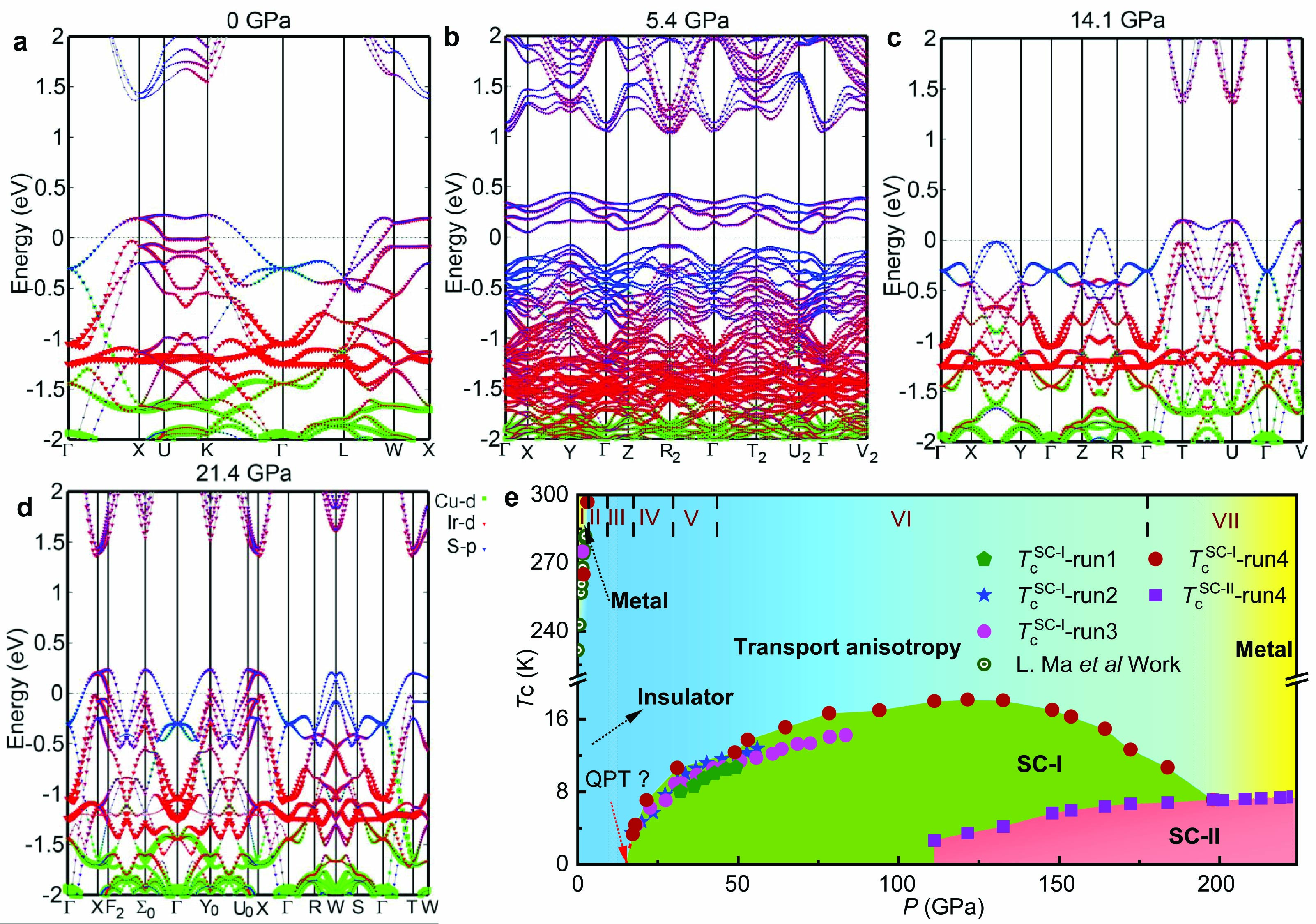}
  \caption{\textbf{Electronic structures and pressure--temperature phase diagram of CuIr$_2$S$_4$.} (\textbf{a}–\textbf{d}) Band structure of CuIr$_2$S$_4$ obtained by DFT calculations under different pressures at (\textbf{a}) 0\,GPa, (\textbf{b}) 5.4\,GPa, (\textbf{c}) 14.1\,GPa and (\textbf{d}) 21.4\,GPa. The orbital characters of bands are represented by different colors, and the projected weights are indicated by marker sizes. (\textbf{e}) Pressure--temperature phase diagram of CuIr$_2$S$_4$. Black dashed lines denote the phase boundaries of structures under pressure at room temperature. Different symbols represent the $T_{\rm MIT}$ and $T_{\rm c}$ of SC‑I and SC‑II measured in different runs. Slight variations in $T_c$ observed between different experimental runs within the SC-I regime can be attributed to differences in diamond culet size, sample dimensions, and pressure calibrations used in each run. Transport anisotropy is determined from the van der Pauw measurements of $R_{12}$ and $R_{14}$ detailed in the SI (Figs.~S6–S7).}
  \label{fig:5}
\end{figure}

To elucidate the physics underlying the observed superconductivity in CuIr$_2$S$_4$ under pressure, we utilized first-principles calculations based on density functional theory (DFT). We investigated the electronic structure of CuIr$_2$S$_4$ across the first four structural phases. Figure~\ref{fig:5}a,b and Fig.~S5 reveal the dominance of Ir-5$d$ states at the Fermi level, with moderate hybridization from S-3$p$ states, while Cu-3$d$ contributions are nearly negligible. At ambient pressure, CuIr$_2$S$_4$ exhibits a metallic ground state, consistent with previous findings~\cite{Sasaki2004Band}. Upon compression, distinct transitions occur: a 0.2--0.3~eV energy gap opens at 5.4~GPa in the triclinic structure. At 14.1~GPa (phase III) and 21.4~GPa (phase IV), CuIr$_2$S$_4$ reverts to metallic ground states. This indicates that structural transitions are pivotal in the emergence of superconductivity. Notably, the bands cross the Fermi surface and exhibit dispersion along $\Gamma$-T direction (close to the [111] crystallographic direction in the spinel phase), while they are relatively flat near the Fermi level along $\Gamma$-X/Y/Z directions in phase III. This is consistent with our anisotropic resistance results in the Supporting Information (SI)~\cite{SM}. We also observe a pronounced in-plane resistance anisotropy in the SC-I phase (see SI~\cite{SM}). Rather than indicating magnetic order, this anisotropy—which vanishes upon entering the zero-resistance SC-II phase—resembles the electronic nematicity or stripe-like order observed in high-$T_c$ cuprates. This suggests that SC-I represents a superconducting state intertwined with a competing order parameter (e.g., nematic or charge density wave). While magnetic fluctuations may contribute to the residual resistance, the robust suppression of the transition under magnetic fields confirms the superconducting nature of this phase. Moreover, in phase IV, where superconductivity first appears, the hybridization between Ir-5$d$ and S-3$p$ orbitals near the Fermi level becomes stronger, suggesting that the sizable $d$--$p$ coupling may play an important role in both metallization and superconductivity.

Combining our electrical transport and structural data, we construct the pressure–temperature ($P$–$T$) phase diagram of CuIr$_2$S$_4$ (Fig.~\ref{fig:5}e). This diagram encapsulates the sequence of pressure-driven phases and transitions. At ambient pressure, the system is metallic at room temperature and undergoes a metal–insulator transition (MIT) into a charge-ordered insulating state below $230$~K. With increasing pressure, the MIT is suppressed and is no longer resolved by $\sim 4$~GPa; beyond this point the low-temperature state remains insulating, but it is reached without a finite-temperature MIT—i.e., the system enters insulating Phase~II. Further compression weakens the insulating behavior in Phase~III and a superconducting phase (SC‑I) emerges in Phase~IV near $18$~GPa, hinting at a nearby quantum critical pressure.

SC‑I appears as a dome-shaped region: $T_c^{\mathrm{SC\mbox{-}I}}$ rises sharply, reaches a maximum of $18.2$~K at approximately $133.3$~GPa, and then decreases with further compression. This behavior contrasts with CuRh$_2$S$_4$, where superconductivity abruptly vanishes near $5.6$~GPa with a reentrance into an insulating state~\cite{Ito2003Pressure}. At higher pressures a second superconducting regime, SC‑II, emerges as a distinct region of the phase diagram. SC‑II appears around $111.8$~GPa and overlaps with SC‑I from roughly $112$–$185$~GPa within Phase~VI; beyond about $185$~GPa, as the system enters Phase~VII, only SC‑II persists, exhibiting $T_c^{\mathrm{SC\mbox{-}II}} \approx 7.5$~K at $224$~GPa. Within the coexistence window, the SC‑II signature increasingly dominates the transport response at higher pressure. The maximum $T_c$ we observe is among the highest reported for spinel superconductors, underscoring frustration relief as an effective design principle~\cite{Moshopoulou1999Superconductivity,Robbins1967Superconductivity,Hagino1995Superconductivity}.

Importantly, the evolution between these two superconducting regimes is continuous rather than abrupt. As pressure increases across the coexistence region, the higher-$T_c$ part of the resistive transition associated with SC-I gradually weakens, while the lower-$T_c$ part associated with SC-II grows, producing a two-step transition that evolves into a single zero-resistance transition at the highest pressures [Figs.~1d–e, 2d–e, 4e]. The closely similar current and field dependences observed in the two regimes, together with this continuous evolution of $T_c$ and the eventual zero resistance, strongly indicate that SC-I and SC-II are two superconducting phases of pressurized CuIr$_2$S$_4$ that originate from the same underlying electronic system, rather than from an impurity phase or a separate antiferromagnetic state.

It is instructive to compare this pressure-tuned superconductivity in CuIr$_2$S$_4$ with Zn-doped CuIr$_2$S$_4$ and with the conventional spinel superconductor CuRh$_2$Se$_4$. Additionally, Zn doping in CuIr$_2$S$_4$ partially suppresses the MIT and induces superconductivity, albeit with relatively low $T_c$ values (2–3 K). However, the Zn-doped phase retains the cubic spinel structure, suggesting a more conventional pairing mechanism similar to CuRh$_2$Se$_4$. In contrast, the high-pressure superconductivity reported here (up to 18 K) emerges in a distinct orthorhombic structure (Phase~IV) and exhibits a dome-shaped, non-monotonic $T_c$ evolution. This comparison suggests that high pressure accesses a different, likely unconventional superconducting state driven by the interplay between frustration relief and lattice distortion, rather than the conventional pairing observed in the chemically disordered cubic phase.

The overall phase diagram thus highlights an intimate interplay between structural transitions, relief of frustration, and the emergence of two superconducting regimes, rather than a single universal mechanism. Superconductivity (SC‑I) first appears only after the lattice departs the cubic Phase~I and enters Phase~IV, emphasizing the crucial role of structure in enabling high-$T_c$ pairing. At ambient conditions CuIr$_2$S$_4$ is a dimerized insulator with Ir$^{4+}$–Ir$^{4+}$ spin-singlet dimers on the geometrically frustrated pyrochlore $B$‑sublattice. Under pressure, our X‑ray diffraction  shows that dimer order destabilizes concomitantly with a structural transition from triclinic to orthorhombic symmetry, supporting a scenario in which pressure relieves frustration‑induced localization and enables itinerant pairing, consistent with predictions for spinels under compression~\cite{Radaelli2005Orbital,khomskii2005orbitally,efthimiopoulos2016structural}.

Within the superconducting state, we observe a large upper critical field $\mu_0 H_{c2}(0)\!\approx\!55$~T at a representative pressure where $T_c\!=\!11.8$~K. This exceeds the Clogston–Chandrasekhar (Pauli) paramagnetic limit $H_P\!\approx\!1.84\,T_c\!\approx\!21.7$~T by a factor of $\sim\!2.53$, and implies a short coherence length
$\xi(0)\!=\!\sqrt{\Phi_0/(2\pi \mu_0 H_{c2}(0))}\!\approx\!2.45$~nm. 
Such a strong Pauli‑limit exceedance points to prominent spin–orbit scattering, strong coupling, multiband effects, and/or proximity to quantum criticality; $T_c(P)$ alone does not fix the mechanism.

The emergence of SC‑II at higher pressures, with lower $T_c$, indicates a distinct superconducting response. It may reflect an additional superconducting channel stabilized by subtle electronic instabilities (e.g., a Lifshitz transition) or a reorganization of Ir orbital states in the high‑symmetry lattice. Alternatively, SC‑I and SC‑II could correspond to electronically or spatially distinct domains enabled by non‑hydrostatic stress at extreme pressures; the extended coexistence window and differences in field response argue for intrinsic distinctions, but complementary bulk probes are needed. A useful conceptual framework is the paired‑electron crystal/pair‑density‑wave scenario~\cite{dayal2011paired}, in which localized singlet pairs—proposed for CuIr$_2$S$_4$ by Khomskii and Mizokawa~\cite{khomskii2005orbitally}—are pressure‑delocalized into a coherent condensate. In this view, geometric frustration, strong correlations, and carrier density near half‑filling cooperate to stabilize superconductivity once dimer order is suppressed.

More broadly, our results show that in a frustrated spinel, superconductivity can emerge directly from an insulating dimer state rather than from a conventional metal. This highlights a route—structural and electronic tuning that relieves frustration—to achieve high-$T_c$ pairing and suggests strategies for engineering quantum materials with intertwined orders. Targeted future measurements—including ac susceptibility under pressure to establish bulk volume fractions, NV‑diamond magnetometry for vortex imaging~\cite{bhattacharyya2024imaging}, and Ir $L$‑edge RIXS to track orbital and spin excitations~\cite{Chen2022Novel,rueff2010inelastic}—will be decisive in determining the pairing mechanism(s) in SC‑I and SC‑II and clarifying their relationship to the collapse of dimer order.

\vspace{3mm}

\emph{Acknowledgments}---
This work was supported by the National Natural Science Foundation of China (NSFC) grant No. 11874075 and 11934017, Science Challenge Project No. TZ2016001, and National Key Research and Development Program of China 2018YFA0305703, 2022YFA1403901 and 2021YFA1400300. The XRD measurements at SPring-8 were carried out under the proposals of 2019A4132. The XRD experiments performed at GeoSoilEnviroCARS (The University of Chicago, Sector 13), Advanced Photon Source (APS), Argonne National Laboratory is supported by the National Science Foundation – Earth Sciences (EAR – 1634415), this research used resources of the Advanced Photon Source, a U.S. Department of Energy (DOE) Office of Science User Facility operated for the DOE Office of Science by Argonne National Laboratory under Contract No. DE-AC02-06CH11357. J.P. Hu acknowledges supports from New Cornerstone Investigator Program and H.-K. Mao acknowledges supports from the National Natural Science Foundation of China Grant No. U1530402 and U1930401. B.J. Chen thanks Dr. X.X.Wu, Dr. Z.Deng, Dr. S.J.Zhang and Dr. S M.Feng for the help during transport experiments and valuable discussions.

\vspace{3mm}


\vspace{3mm}


\vspace{3mm}

\bibliography{full_references_abbreviated}

\newpage

\end{document}